\documentclass[dvips]{acta}
\usepackage{supertabular,lscape,epsfig}
\usepackage{amssymb}
\usepackage{amsmath}

\SetPages{0}{0}

\SetVol{76}{2026}

\newcommand\fs{\mbox{$.\!\!^{\mathrm s}$}}% 
\newcommand\arcdeg{\mbox{$^\circ$}}% 
\newcommand\arcmin{\mbox{$^\prime$}}% 
\newcommand\farcs{\mbox{$.\!\!^{\prime\prime}$}}% 

\newcommand\eventK{KMT-2024-BLG-0816}
\newcommand\eventO{OGLE-2024-BLG-0519}
\newcommand\event{\eventK/\eventO}

\begin{document}

\begin{Titlepage}
\Title{KMT-2024-BLG-0816/OGLE-2024-BLG-0519 -- A Microlensing Event with Candidate Free-Floating Planet Lens and Blended Light}

\Author{
R.~~P~o~l~e~s~k~i$^1$,~~ % 0000-0002-9245-6368
Y.-H.~~R~y~u$^2$,~~ % 0000-0001-9823-2907
A.~~U~d~a~l~s~k~i$^1$,~~ % 0000-0001-5207-5619
W.~~Z~a~n~g$^3$,~~\\ % 0000-0001-6000-3463
~~and~~\\
M.\,D.~~A~l~b~r~o~w$^4$,~~ % 0000-0003-3316-4012
S.-J.~~C~h~u~n~g$^2$,~~ % 0000-0001-6285-4528
A.~~G~o~u~l~d$^{5,6}$,~~
C.~~H~a~n$^7$,\\ % 0000-0002-2641-9964
K.-H.~~H~w~a~n~g$^2$,~~ % 0000-0002-9241-4117
Y.\,K.~~J~u~n~g$^{2,8}$,~~ % 0000-0002-0314-6000
I.-G.~~S~h~i~n$^{3}$,~~ % 0000-0002-4355-9838
Y.~~S~h~v~a~r~t~z~v~a~l~d$^{10}$,\\ % 0000-0003-1525-5041
J.\,C.~~Y~e~e$^9$,~~ % 0000-0001-9481-7123
H.~~Y~a~n~g$^{3,11}$,~~ % 0000-0003-0626-8465
D.-J.~~K~i~m$^2$,~~
C.-U.~~L~e~e$^2$,~~ % 0000-0003-0043-3925
B.-G.~~P~a~r~k$^2$,\\ % 0000-0002-6982-7722
(KMTNet Collaboration)\\
~~and~~\\
M.\,K.~~S~z~y~m~a~ń~s~k~i$^1$,~~ % 0000-0002-0548-8995
I.~~S~o~s~z~y~ń~s~k~i$^1$,~~ % 0000-0002-7777-0842
K.~~U~l~a~c~z~y~k$^{12,1}$,~~ % 0000-0001-6364-408X
P.~~P~i~e~t~r~u~k~o~w~i~c~z$^1$,\\ % 0000-0002-2339-5899
J.~~S~k~o~w~r~o~n$^1$,~~ % 0000-0002-2335-1730
D.~~S~k~o~w~r~o~n$^1$,~~ % 0000-0001-9439-604X
P.~~M~r~ó~z$^1$,~~ % 0000-0001-7016-1692
K.~~R~y~b~i~c~k~i$^{10,1}$,~~ % 0000-0002-9326-9329
P.~~I~w~a~n~e~k$^1$,\\ % 0000-0002-6212-7221
M.~~W~r~o~n~a$^{13,1}$,~~ % 0000-0002-3051-274X
M.~~G~r~o~m~a~d~z~k~i$^1$,~~ % 0000-0002-1650-1518
M.~~M~r~ó~z$^1$,~~ % 0000-0002-8911-6581
M.~~R~a~t~a~j~c~z~a~k$^1$\\ % 0000-0002-3218-2684
(OGLE Collaboration)}{%~} 
%}{}
${}^1$ Astronomical Observatory, University of Warsaw, Al. Ujazdowskie 4,\\ 00-478 Warszawa, Poland \\
${}^2$ Korea Astronomy and Space Science Institute, Daejeon 34055, Republic of Korea\\
${}^3$ Department of Astronomy, Westlake University, Hangzhou 310030, Zhejiang Province, China\\
${}^4$ University of Canterbury, School of Physical and Chemical Sciences, Private Bag 4800, Christchurch 8020, New Zealand\\
${}^5$ Max-Planck-Institute for Astronomy, K\"onigstuhl 17, 69117 Heidelberg, Germany\\
${}^6$ Department of Astronomy, Ohio State University, 140 W. 18th Ave., Columbus,\\ OH 43210, USA\\
${}^7$ Department of Physics, Chungbuk National University, Cheongju 28644,\\ Republic of Korea\\
${}^8$ National University of Science and Technology (UST), Daejeon 34113,\\ Republic of Korea\\
${}^9$ Center for Astrophysics $|$ Harvard \& Smithsonian, 60 Garden St.,\\ Cambridge, MA 02138, USA\\
${}^{10}$ Department of Particle Physics and Astrophysics, Weizmann Institute of Science, Rehovot 7610001, Israel\\
${}^{11}$ Department of Astronomy, Tsinghua University, Beijing 100084, China\\
${}^{12}$ Department of Physics, University of Warwick, Coventry CV4 7AL, UK \\
${}^{13}$ Department of Astrophysics and Planetary Sciences, Villanova University,\\ 800 Lancaster Avenue, Villanova, PA 19085, USA\\
e-mail: rpoleski@astrouw.edu.pl, yhryu@kasi.re.kr}

\Received{Month Day, Year}
\end{Titlepage}

\Abstract{
We present the discovery of a free-floating planet microlensing event KMT-2024-BLG-0816. The event shows finite-source effect, significant blending light, and no microlensing signal from a putative planet host. Among the free-floating planet events with finite source effects, this is the only event with unresolved blending light. We discuss how follow-up observations can be used to determine whether the blending light originates from a putative planet host.
}{
key words: 
gravitational microlensing exoplanet detection --- 
free floating planets --- 
photometry
}

\section{Introduction} % -----------------------------------------------------

The ongoing high-cadence microlensing surveys (Bond \etal 2001; Udalski \etal 2015; Kim \etal 2016) are discovering non-repeating short-timescale events (see Mróz and Poleski 2024 for a recent review). For these events, we measure the Einstein timescales and, if the finite-source effect is detected, the Einstein ring radii $\theta_\mathrm{E}$. Both these quantities are proportional to the square root of the lens mass, and the measured values indicate planetary--mass lenses. Hence, the lenses are naturally interpreted as planets. An alternative interpretation was proposed: planetary--mass primordial black holes (Niikura \etal 2019), which was refuted by further observations (Mróz \etal 2024b).

Once we know the short timescale events are caused by planets, we may ask if these planets are bound to stars or not. The lightcurves of these events do not show signs of other lensing bodies, hence, they are called free-floating planets (FFPs). It is possible that these planets are bound and their hosts are not detected due to large angular separations between the hosts and the sources of the microlesning events. The bound-planet interpretation can be proved in three ways: detailed analysis of the planet peak, detection of a microlensing signal (photometric or astrometric) caused by the host, or detection of the host light (Han \etal 2005; Di Stefano 2012; Ryu \etal 2013, 2016). The latter way is difficult to apply because only one out of the ten previously published FFPs with finite-source effect showed statistically significant blending light (OGLE-2023-BLG-0524, see discussion below; Kapusta \etal 2025)\footnote{The others were presented by: Mróz \etal (2018, 2019, 2020a,b), Kim \etal (2021), Ryu \etal (2021a), Koshimoto \etal (2023), and Jung \etal (2024).}. 
Accurate determination of blending flux is hampered by short length of these events and intrinsic degeneracies (Johnson \etal 2022). 
Additionally, one can place statistical constraints on frequency of hosts by comparing the occurrence rates of the wide-orbit bound planets (Suzuki \etal 2016; Poleski \etal 2021; Zang \etal 2025) and the FFP events (Mróz \etal 2017; Gould \etal 2022; Sumi \etal 2023).

Here we report the discovery of an FFP event \event. The microlensing model fit shows significant blending flux. The blending flux may originate from the lens system, from a companion to the source, or from an unrelated star. If the first origin was the case, then the presented event would be the first FFP for which a host star was detected. 
There are only two published studies in which hosts of FFPs were searched for using high angular resolution imaging. First, five FFPs without finite-source effect were imaged using adaptive optics on the Keck Telescope by Mróz \etal (2024a).
Second, the FFP event OGLE-2023-BLG-0524 had serendipitous archival imaging by the \textit{Hubble Space Telescope} taken 26 years before the event (Kapusta \etal 2025). The fitted microlensing model indicated that the blend-to-source flux ratio is $2.33\pm0.65$ in the OGLE data. The archival \textit{Hubble} image revealed three stars not resolved in the seeing-limited ground-based images. These stars are of similar brightness, and a recent \textit{Hubble} image clearly showed that the two neighboring stars are not the hosts of the planet. The target star is slightly brighter than the source flux derived from the microlensing model but the difference is not statistically significant. 
Neither Mróz \etal (2024a), nor Kapusta \etal (2025) have detected planet hosts, however, they also have not fully probed host distance and luminosity parameter space.

We note that it is routinely checked whether short single-peaked events show microlensing signals of binary lenses (see Section 7.1 of Ryu \etal 2021b). An illustrative example for that is a single-peaked event with $t_\mathrm{E} = 3.54\pm0.05~\mathrm{d}$ in which a slight peak asymmetry revealed the binary nature of the lens (Han \etal 2020).

This paper is structured as follows. The following section presents photometric data used. Their analysis is presented in Sections~3. Sections~4 and 5 discuss the color-magnitude diagram (CMD) and astrometry, respectively. In Section~6 we reanalise the event assuming there is no blending light in the OGLE data. We end with a discussion that includes prospects for follow-up observations.

\section{Photometric Data} % -----------------------------------------------------

A microlensing event located at $\mathrm{RA}=17^{\rm h}58^{\rm m}36\fs 145$, $\mathrm{Dec.}={-29}\arcdeg{23}\arcmin{26}\farcs{33}$ (J2000; $(l, b) = (1.05\arcdeg, -2.67\arcdeg)$) was announced on May 3rd, 2024 (JD: 2460434). 
The first alert was announced by the Korean Microlensing Telescope Network (KMTNet) Event Finder (Kim \etal 2018) as \eventK. On the same day, the Optical Gravitational Lensing Experiment (OGLE) survey independently announced this event as \eventO via its Early Warning System (Udalski 2003). 

The KMTNet uses three identical telescopes widely spread in longitude in order to allow continuous monitoring of the sky (Kim \etal 2016). 
The telescopes are located at 
the Cerro-Tololo Inter-American Observatory (CTIO) in Chile, 
the South African Astronomical Observatory (SAAO) in South Africa, 
and the Siding Spring Observatory (SSO) in Australia. 
Each telescope has 1.6 m diameter and is equipped with a CCD camera. The camera field of view is 4 deg$^2$ and the pixel scale is 0.4 arcsec. The event falls in two overlapping fields called BLG02 and BLG42. The data from each field are analyzed separately. The KMTNet data were taken with a cadence of 17\,min. (SAAO and SSO) and 24\,min. (CTIO) in the $I$-band. For the $V$-band data the cadence was 100 and 150\,min., respectively. We analyze six $I$-band datasets totaling 4985 epochs and four $V$-band (from CTIO and SSO) datasets totaling 339 epochs.

The OGLE survey uses a single 1.3 m telescope located at the Las Campanas Observatory in Chile (Udalski \etal 2015). The telescope is equipped with a camera that provides a field of view of 1.4 deg$^2$ and a pixel scale of 0.26 arcsec. The target star was observed in the $I$ band with a cadence of 1\,hr (582 epochs are analyzed here). The same target was observed in the earlier phases of the OGLE project that used the same telescope and smaller FoV cameras but the same pixel scale: 
OGLE-II (329 $I$-band epochs between Apr 1997 and Oct 2000) and 
OGLE-III (1304 $I$-band epochs between Jul 2001 and May 2009). 
The OGLE-II and OGLE-III photometry does not show any signs of variability. There is only a single OGLE $V$-band epoch taken during the event, so we do not analyze $V$-band photometry from OGLE-IV.

We extract the photometry using the difference image analysis (Alard and Lupton 1998; Woźniak 2000). The KMTNet photometric pipeline is implemented based on Albrow \etal (2009), Yang \etal (2024), and Yang \etal (2025). The OGLE pipeline was presented by Udalski \etal (2015). The difference image analysis typically underestimates photometric uncertainties, hence, we correct the uncertainties. We fit a preliminary model and find that four of the KMTNet $I$-band datasets exhibit $\chi^2/\mathrm{dof}$ significantly larger than one. In these cases, we multiply uncertainties by constants that result in $\chi^2/\mathrm{dof} \approx 1$: 
CTIO BLG02 -- $1.40$, 
CTIO BLG42 -- $1.23$, 
SSO BLG02 -- $1.16$, and
SAAO BLG02 -- $1.06$. 
For the OGLE data, we scaled the uncertainties using prescriptions of Skowron \etal (2016).

In addition to the time series photometry presented above, we also used a NIR photometric catalog in order to measure the NIR extinction. For this purpose, we use the VVV Infrared Astrometric Catalogue version 2 (VIRAC2; Smith \etal 2025). From this catalog we use only the $J$ and $K_s$ (indicated here as $K$) band data. The VIRAC2 catalogue is based on the data collected by the 4 m Visible and Infrared Survey Telescope (VISTA) at the Cerro Paranal Observatory in Chile.

\section{Light Curve Analysis} % -----------------------------------------------------

The event light curve shows a single short-lasting peak (Fig.~1). The rising and falling branches of the peak are steep, indicating that finite source effect is important. 

\begin{figure}[htb]
\includegraphics[width=\textwidth]{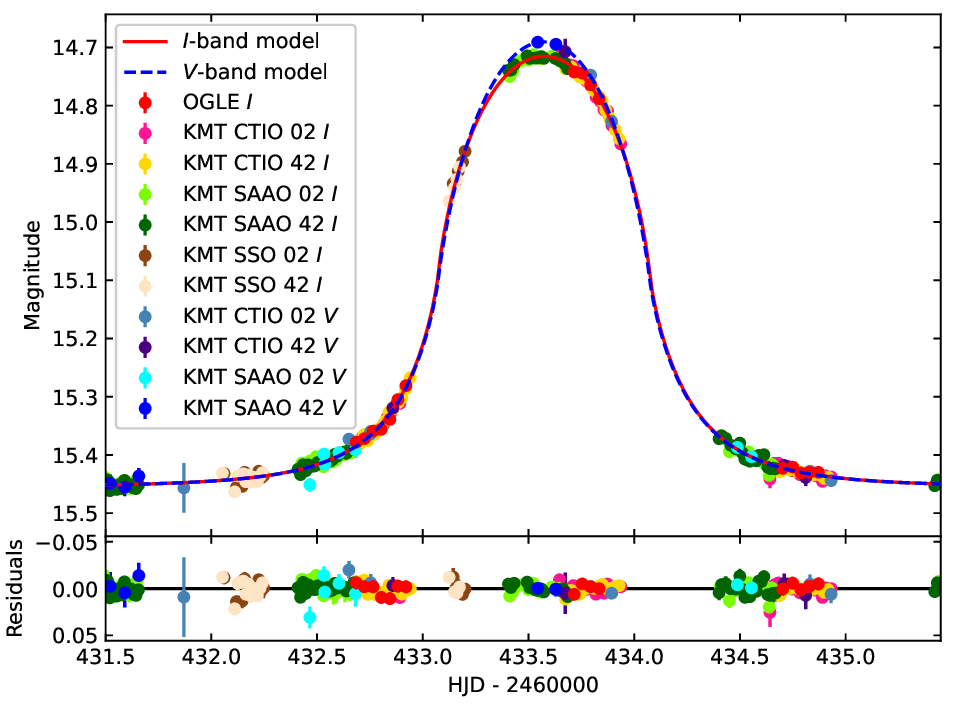}
\FigCap{Light curve of \event. The KMTNet data are shown separately for each observatory, field (02 or 42), and band.}
\end{figure}

\subsection{Single-lens Model} % -----------------------------------------------------

We parameterize the microlensing model using the following parameters:
$t_0$ -- the epoch of minimum lens-source separation,
$u_0$ -- the signed impact parameter relative to $\theta_\mathrm{E}$,
$t_\mathrm{E}$ -- the Einstein timescale, and 
$t_\star \equiv \rho t_\mathrm{E}$ -- the source radius crossing time;
$\rho$ is the source radius relative to $\theta_\mathrm{E}$.  
We also fit for two flux parameters per dataset: the source flux $f_s$ and the blending flux $f_b$. For each set of $(t_0, u_0, t_\mathrm{E}, t_\star)$ we fit for flux parameters via linear regression. We use \textsc{MulensModel} software (Poleski and Yee 2019) for fitting and call pre-computed tables (Bozza \etal 2018) 
to evaluate the magnification of a finite-source with limb darkening and a single lens. 
For posterior sampling we use the ensemble Monte-Carlo Markov chain sampler implemented by Foreman-Mackey \etal (2013; \textsc{emcee}). 

We use a linear limb-darkening law with coefficients calculated based on the source properties in a preliminary fitted model. 
The source is close to the Red Clump (RC) on the CMD (see below), and the measured reddening-corrected color $(V-I)_0$ corresponds to an effective temperature of $\approx 5080$\,K (Houdashelt \etal 2000). 
We adopt this value for the effective temperature, adopt the typical surface gravity of the RC stars $\log g = 2.5$, as well as the typical metallicity for the bulge stars $\mathrm{[Fe/H]} = -0.5$ (Barbuy \etal 2018). We interpolate the linear limb-darkening coefficients of Claret and Bloemen (2011) to these values and obtain $u_I = 0.52$ and $u_V = 0.70$.  
After the final model fitting, we verify that it resulted in parameters of the source that are consistent with the preliminary fitted model. 

The best-fitted model is shown in Fig.~1. We present the parameter distribution in Fig.~2 and Tab.~1. The table provides the blend-to-source flux ratios ($f_b/f_s$) for the $I$ and $V$ bands. Figure~2 shows the posterior for four fitted parameters and additionally source brightness in the OGLE $I$ band. We find that $0.18\%$ of samples have negative blending flux in the OGLE $I$ band.

\begin{figure}[t]
\includegraphics[width=\textwidth]{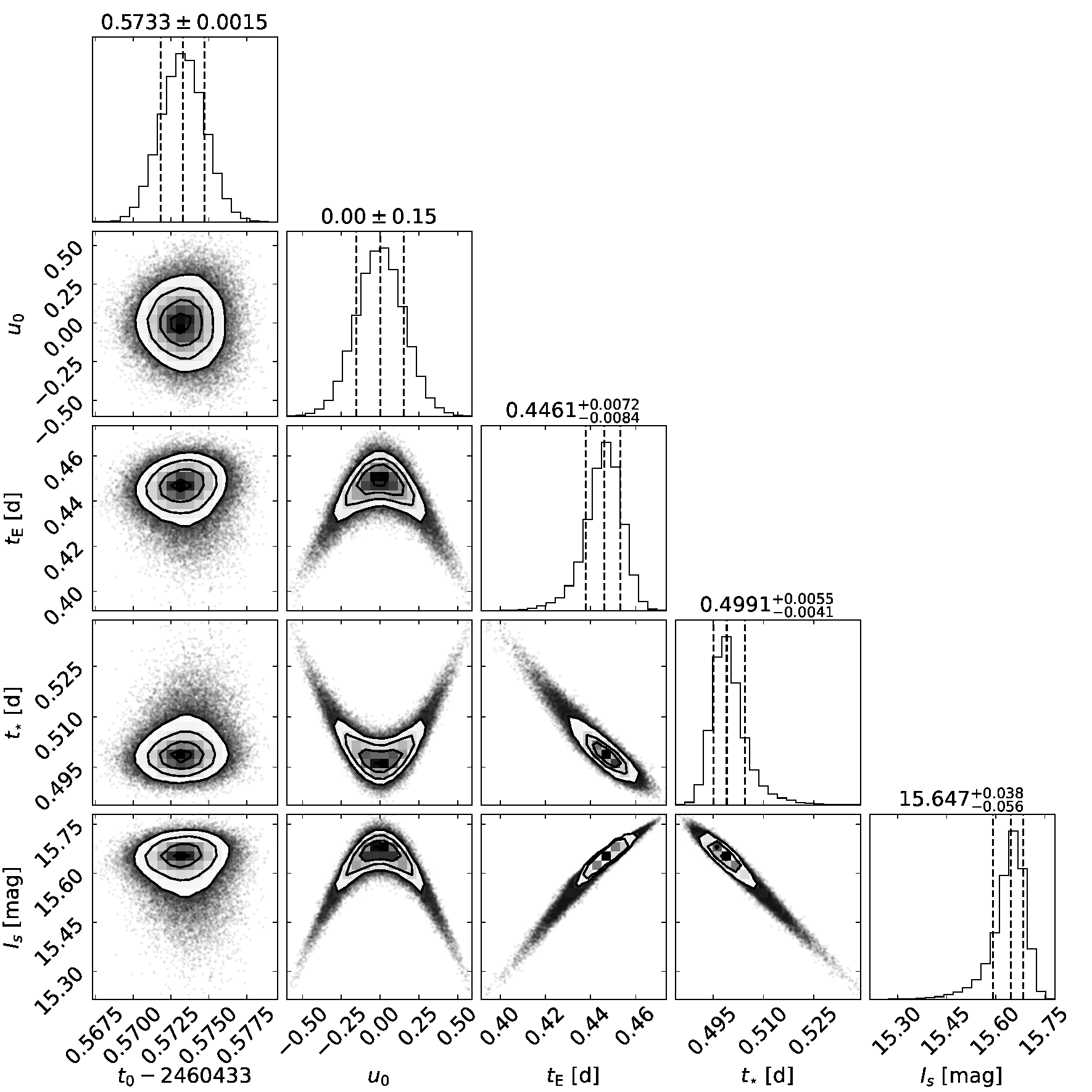}
\FigCap{Posterior distribution of single lens parameters.}
\end{figure}

\MakeTable{l|r}{6.5cm}{Parameters of single-lens finite-source model fitted}
{\hline
parameter &  value\\
\hline
$t_0$ & $2460433.5733\pm0.0015$ \\
$u_0$ & $0.00\pm0.15$ \\
$t_\mathrm{E}$ (d) & $0.4461^{+0.0072}_{-0.0084}$ \\
$t_\star$ (d) & $0.4991^{+0.0055}_{-0.0041}$ \\
\hline
$\rho$ & $1.119^{+0.033}_{-0.026}$ \\
\hline
$f_{b, I}/f_{s, I}$ & $0.195^{+0.044}_{-0.051}$ \\
$f_{b, V}/f_{s, V}$ & $0.113^{+0.037}_{-0.047}$ \\
\hline
$\chi^2/\mathrm{dof}$ & 5401.84/5880 \\
\hline
\multicolumn{2}{p{4.5cm}}{First four rows show statistics of fitted parameters.}
}

\subsection{Binary-lens Model} % -----------------------------------------------------

For binary-lens models, we use the \textsc{VBBL} software (Bozza \etal 2018) to evaluate magnification. These calculations require a root solver for the fifth order complex polynomial for which the Skowron and Gould (2012) algorithm is used. We fit models starting from a large grid of parameters: 
\begin{itemize}
	\item 21 values of $s$ from 1 to 10 with 0.05 spacing in $\log{s}$,
	\item 21 values of $q$ from $10^{-4.5}$ to $10^{-0.5}$ with 0.2 spacing in $\log{q}$,
	\item 19 values of $\alpha$ from $0^{\circ}$ to $180^{\circ}$ with $10^{\circ}$ spacing.
\end{itemize}
For comparison, we note that the widest-orbit microlensing planet has $s=5.26\pm0.11$ and $q=(2.41\pm0.45)\times10^{-4}$ (Poleski \etal 2014). We estimate the starting parameters of the binary-lens models based on the single-lens results and assumed $(s, q, \alpha)$:
$$t^\mathrm{2L}_\mathrm{E} = \sqrt{\frac{1+q}{q}} t_\mathrm{E} $$
$$u^\mathrm{2L}_0 = u_0\sqrt{\frac{q}{1+q}} + \frac{\sin\alpha}{1+q}\left(s - \frac{1}{s}\right)$$
$$t^\mathrm{2L}_0 = t_0 + \frac{t^\mathrm{2L}_\mathrm{E}\cos\alpha}{1+q}\left(s - \frac{1}{s}\right)$$
We perform the fits in three stages. First, we fit binary-lens models with $(s, q, \alpha)$ fixed and ignoring limb darkening (which speeds-up calculations). Second, we re-run the fits with the starting parameters taken from the best models in the first stage and include limb-darkening (the same coefficients as for the single lens). Third, we continue the runs with all binary-lens parameters freed. The best solution we find has $\chi^2 = 5382.62$ ($s=7.13\pm0.15$ and $q=0.0204\pm0.0040$). The improvement of $\chi^2$ when compared to the single lens model is 19.22, which is too low to reject the single-lens interpretation.

Next, we establish limits on a putative host of the planet. We follow the methodology presented by Gaudi and Sackett (2000). We start with the results of the second step described above, i.e., obtained with fixed $(s,q,\alpha)$. We notice that some of the fitted models resulted in a significantly negative $f_{b,I}$. We repeat these runs of the grid this time rejecting all models with $f_{b,I}$ more negative than corresponding to a ``negative'' $20~\mathrm{mag}$ star. We are not aware of discussion of possibly negative blending flux in previous papers presenting host detection efficiency calculations for FFPs, hence, the models with significantly negative blending flux may have been accepted. 

We apply the host detection criterion $\Delta\chi^2 > 25$ and present the results as a function of $s$ and $Q=1/q$, following Jung \etal (2024). The host detection sensitivity is presented in Fig.~3. We note that models with grid points with the smallest values of $s$ resulted in models with $t_\mathrm{E} < 1$, i.e., systems composed most likely of two planets. Calculation of host detection efficiency for FFP events that takes into account priors on the host mass, priors on the Galactic model (both affect $t_\mathrm{E}$), and negative blending flux remains to be done. 

\begin{figure}[htb]
\includegraphics[width=\textwidth]{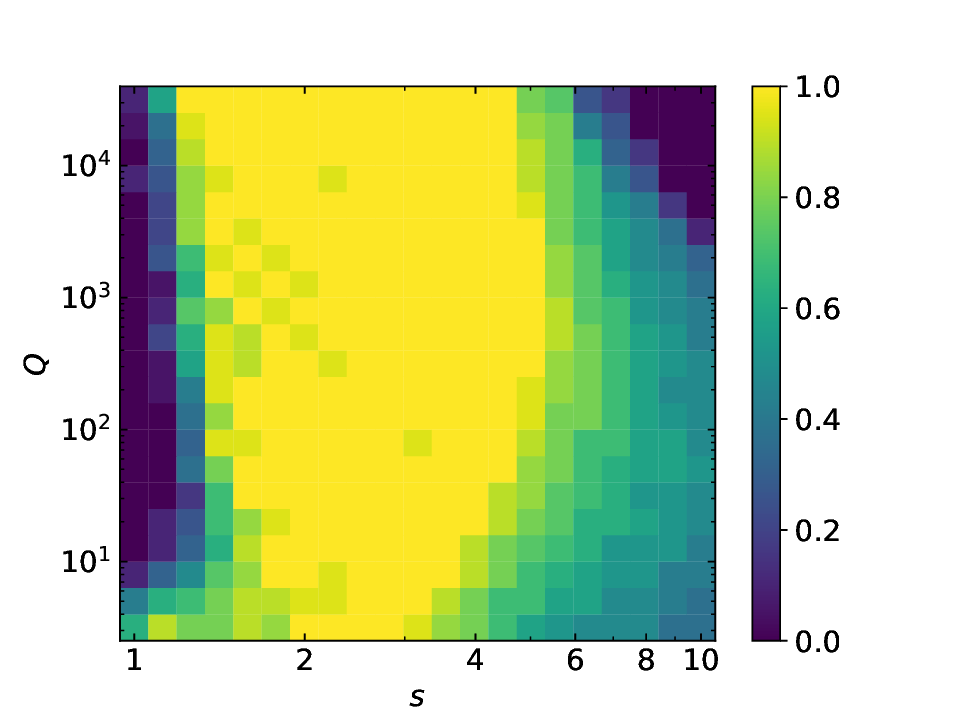}
\FigCap{Host detection efficiency as a function of projected separation in units of $\theta_\mathrm{E}$ and the host-to-star mass ratio.}
\end{figure}

\section{Color-magnitude Diagram} % -----------------------------------------------------

We derive the source properties using the method originally proposed by Yoo \etal (2004). The first step in the analysis of the CMD is the calibration of all magnitudes to a common photometric system. For the $I$-band, we use the OGLE-IV photometry. However, we cannot analyze source properties based on the OGLE $V$-band photometry because there is only a single measurement during the magnified part of the light curve. Among the available KMT $V$-band datasets we select CTIO BLG02 and calibrate the zero-point of this dataset magnitude scale using nearby stars that are common with the OGLE-III photometric maps (Szymański \etal 2011). This calibration is precise to $0.010~\mathrm{mag}$. OGLE-III was used instead of OGLE-IV because in the latter the target star is closer to the edge of a CCD chip. We note that the OGLE-IV photometry was calibrated to the OGLE-III scale (Udalski \etal 2015). Finally, we obtain calibrated baseline brightness: 
$$I_\mathrm{base} = 15.455~\mathrm{mag},
\qquad
V_\mathrm{base} = 17.494~\mathrm{mag}.$$
In parallel, we use VIRAC2 to measure the $K$-band extinction that is needed to plan follow-up observations. We show the optical CMD in Fig.~4.

\begin{figure}[htb]
\includegraphics[width=\textwidth]{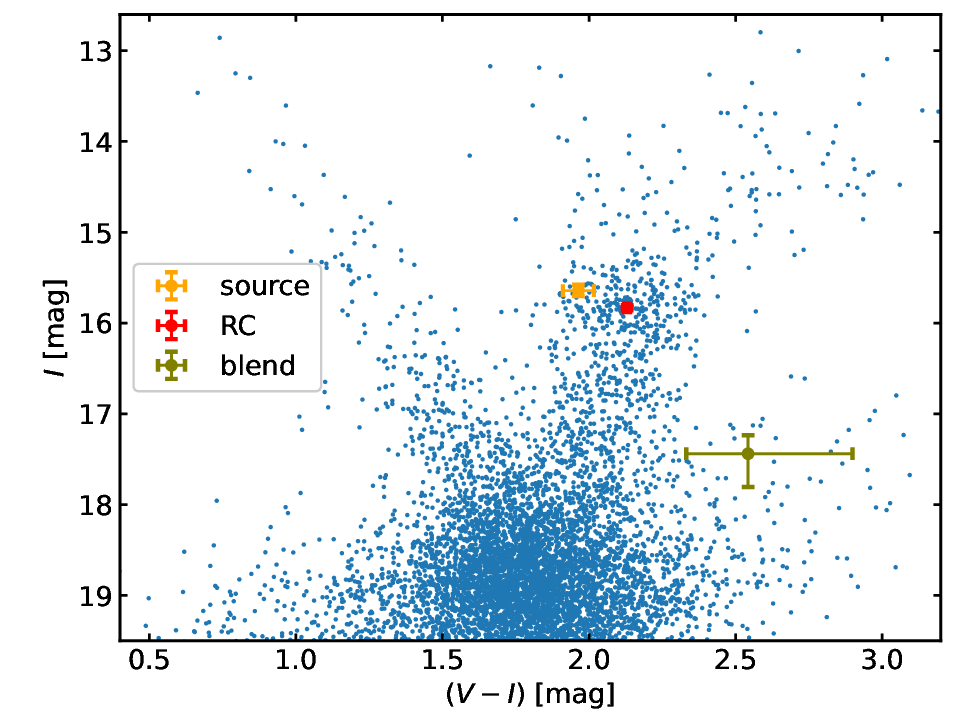}
\FigCap{Color-magnitude diagram of stars within $1.5\arcmin$ of the event.}
\end{figure}

The next step of the analysis is to calculate the reddening-corrected brightness of the source. This calculation is based on the source flux estimated during microlensing model fitting. In order to fully account for parameter correlations in posterior samples, we use the full posterior, not just point estimates. We additionally account for uncertainties in calibration of the procedure described below by adding gaussian noise with zero mean and standard deviation of $0.05~\mathrm{mag}$ (Bensby \etal 2013; Gould \etal 2014).

We derive the reddening and extinction by comparing the observed and reddening-free properties of the RC. The latter are assumed to be:
$$I_{\mathrm{RC},0} = 14.395~\mathrm{mag},$$
$$(V-I)_{\mathrm{RC},0} = 1.06~\mathrm{mag},$$ 
$$K_{\mathrm{RC},0} = I_{\mathrm{RC},0} - M_{I,\mathrm{RC}} + M_{K,\mathrm{RC}} = 13.015~\mathrm{mag}$$
(Bensby \etal 2013; Nataf \etal 2013). We assume the uncertainty of $(V-I)_{\mathrm{RC},0}$ to be $0.02~\mathrm{mag}$ based on Bensby \etal (2017). The observed RC properties are derived following the Nataf \etal (2013) method and using stars within a radius of $1.5\arcmin$ around the target. We obtain:
$$A_I = 1.436\pm0.029~\mathrm{mag},$$ 
$$E(V-I) = 1.070\pm0.010~\mathrm{mag},$$
$$A_K = 0.214\pm0.030~\mathrm{mag}.$$ 
The first two values provide reddening-corrected source properties: 
$$I_{s,0} = 14.212\pm0.050~\mathrm{mag}, 
\qquad
(V-I)_{s,0} = 0.894\pm0.025~\mathrm{mag},$$ 
which we further transform to the $K$ band using calibrations from Bessell and Brett (1988):
$$K_{s,0} = 13.102\pm0.072~\mathrm{mag},
\qquad
(V-K)_{s,0} = 2.002\pm0.076~\mathrm{mag}.$$
We apply angular diameter-color relation presented by Adams \etal (2018) and obtain the normalized source radius:
$$\theta_\star = 5.53\pm0.22~\mathrm{\mu as}.$$  
This value is then used to calculate the Einstein ring radius and the relative lens-source proper motion:
$$\theta_\mathrm{E} = \theta_\star/\rho = 4.95\pm0.18~\mathrm{\mu as},
\qquad
\mu_\mathrm{rel} =  4.05\pm0.15~\mathrm{mas/yr}.$$
The $\theta_\mathrm{E}$ values measured for single lenses show a dearth of values between 10 and $30~\mathrm{\mu as}$ (Ryu \etal 2021a; Gould \etal 2022). The value measured for {\event} is typical for objects below the lower boundary, i.e., FFPs (Mróz and Poleski 2024).

The CMD position of the blend is consistent within $1.5\sigma$ with the bulge giant branch. The other possibility is that the blend is a main sequence star. Its de-reddened color suggests a spectral type of approximately K4. Based on the absolute brightness of such stars, we estimate the blend distance in this scenario to be on the order of $1.2~\mathrm{kpc}$.

In order to verify the interpretation of the blending flux as coming from an unrelated star, we calculate the density of stars with $I$-band brightness between 17 and $18~\mathrm{mag}$ in the vicinity of the event. We find 0.035 stars per arcsec$^2$. The reference image has a FWHM of $0\farcs 81$, and the expected number of stars in the above brightness range in a circle with the radius of the reference image FWHM is $0.71$.

\begin{figure}[t]
\includegraphics[width=\textwidth]{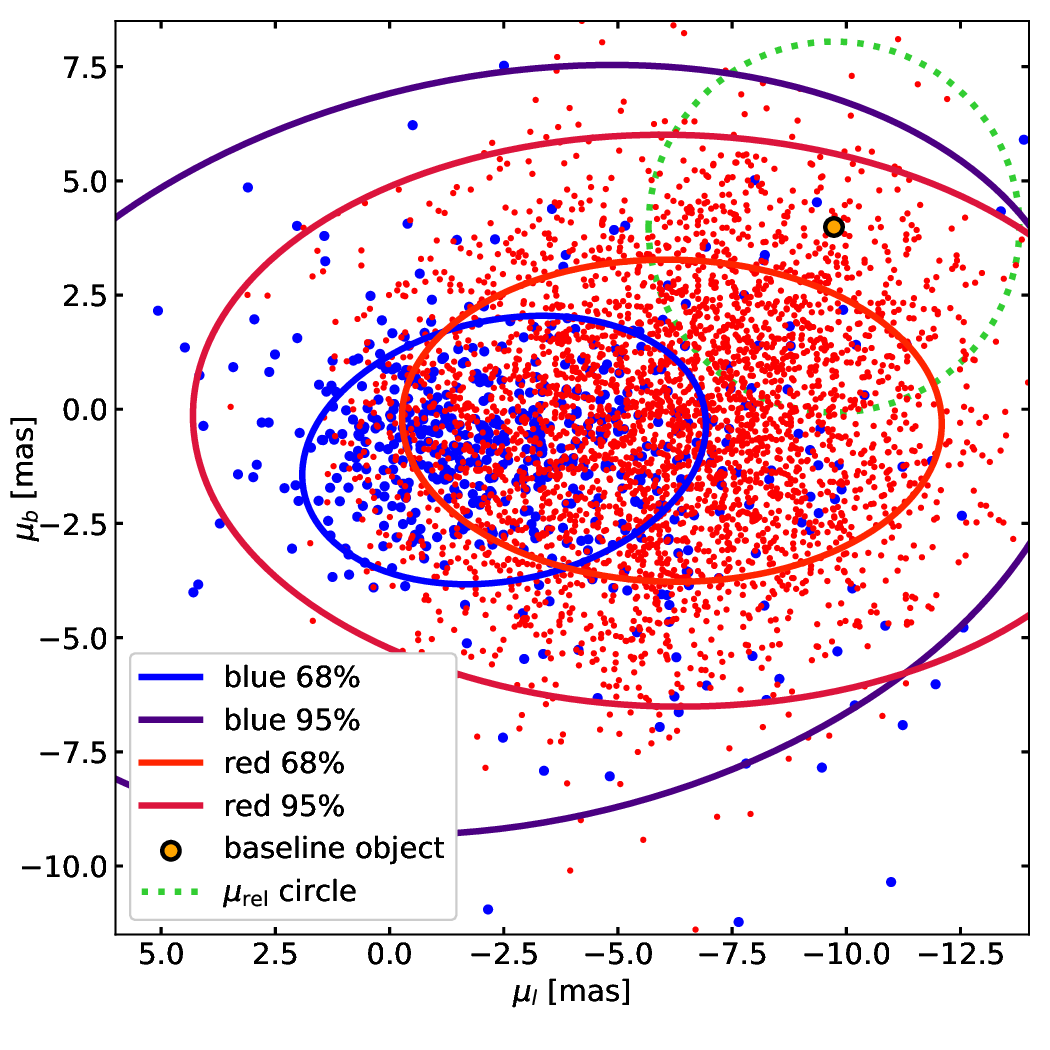}
\FigCap{Galactic frame proper motion vector-point diagram for stars close to the event. We show blue (disk) and red (bulge stars). In each case we also plot $68\%$ and $95\%$ ellipses. The proper motion of the baseline object (i.e., flux-weighted proper motion of source and blend) is indicated by an orange dot and a green dotted circle around it has radius of $\mu_\textrm{rel}$. The lens proper motion should be close to this circle.}
\end{figure}

\section{Astrometry} % -----------------------------------------------------

The target star is present in the \textit{Gaia} Data Release 3 (DR3; Gaia Collaboration \etal 2016, 2023) as 4062325478979473024. 
The astrometric parameters (corresponding to the flux-weighted source and blend) of this star are:
$$\mu_\mathrm{R.A.} = -8.303\pm0.091~\mathrm{mas/yr},$$ 
$$\mu_\mathrm{Dec.} = -6.454\pm0.064~\mathrm{mas/yr},$$ 
$$\pi = 0.190\pm0.085~\mathrm{mas}.$$ 
The fit to the \textit{Gaia} astrometric measurements has Renormalized Unit Weight Error (RUWE) of 1.011, which indicates a reliable measurement. 

We also measure proper motion using OGLE time-series astrometry (OGLE--URANUS; Udalski \etal in prep.). We extract centroids from OGLE images and then find the transformation from pixel coordinates to equatorial coordinates using stars that are measured by both OGLE and \textit{Gaia} DR3. The equatorial coordinates of \textit{Gaia} stars are calculated for each epoch of OGLE images using astrometric parameters from the \textit{Gaia} DR3. Then the transformation is applied to all OGLE centroids, resulting in equatorial coordinates tied to the Gaia reference frame. For the target star, we use 7,000 epochs collected over 13.4 years. We obtain:
$$\mu^\mathrm{OGLE}_\mathrm{R.A.} = -8.357\pm0.052~\mathrm{mas/yr},$$
$$\mu^\mathrm{OGLE}_\mathrm{Dec.} = -7.113\pm0.052~\mathrm{mas/yr}.$$
Additionally, we check for potential bias in these measurements using the bootstrap method. We find 
$\sigma^\mathrm{bias}_\mathrm{R.A.} = 0.027~\mathrm{mas/yr}$ and 
$\sigma^\mathrm{bias}_\mathrm{Dec.} = 0.040~\mathrm{mas/yr}$. 
The $\mu_\mathrm{R.A.}$ measurements of \textit{Gaia} and OGLE are consistent. The difference in $\mu_\mathrm{Dec.}$ is small but statistically significant: $0.659\pm0.092~\mathrm{mas/yr}$ or $7.2\sigma$. In the case of blended target, we expect a small difference in measured proper motions because of different blend-to-source flux ratios in $G$ and $I$ bands.

We present the proper motion of the baseline object in comparison with other nearby (within $5'$), bright ($G < 17.5~\mathrm{mag}$), and reliably measured ($RUWE<1.4$) stars in Fig.~5. The stars were divided into blue sample ($(BP-RP)<1.75~\mathrm{mag}$; i.e., stars in the disk) and red sample ($(BP-RP)>1.75~\mathrm{mag}$; i.e., stars in the bulge). 
We find the ellipse containing $68\%$ of the blue and red samples. Then we expand these ellipses so that they contain the baseline object. These expanded ellipses contain $91\%$ of the blue sample and $88\%$ of the red sample. Hence, the baseline proper motion is consistent with both blue and red samples. Taking into account the number of points in each sample, the red sample is preferred, i.e., the source (which dominates the baseline flux) is more likely in the bulge than in the disk. We do not know the direction of $\mu_\mathrm{rel}$, we only know its length. Based on the distribution of points in Fig.~5, we deduce that the lens is more likely located in the bulge, but the disk location is not excluded.

The \textit{Gaia} parallax differs from the bulge parallax of $1/8.3~\mathrm{kpc} = 0.120~\mathrm{mas}$ by only $0.8\sigma$, thus, the parallax measurement is consistent with the combined light of two bulge stars. On the other hand, even if the blend light originates very close to us, e.g., at a distance of 1.5~kpc, when combined with a nine-times brighter source at the bulge distance gives a weighted parallax of $0.1/1.5~\mathrm{kpc}+0.9/8.3~\mathrm{kpc} = 0.175~\mathrm{mas}$, which is very close to the mean value reported by \textit{Gaia}. Hence, the \textit{Gaia} parallax measurement does not constrain the blending light distance.

OGLE data provide not only the time-series astrometry analyzed above but also the reference image and co-added difference image (which provides the most accurate information on the source position). One can compare the centroids measured on these images in order to constrain the blending light. The magnified centroid measured on the co-added difference image is shifted relative to the baseline object by 0.110~pix toward West and 0.060~pix toward South. The magnified centroid is measured with uncertainties of $\sigma_\mathrm{R.A.} = 0.035~\mathrm{pix}$ and $\sigma_\mathrm{Dec.} = 0.022~\mathrm{pix}$, while the uncertainty in the baseline centroid is $\sigma = 0.050~\mathrm{pix}$ in each direction. Hence, R.A. shift is significant at $1.8\sigma$ level and we cannot rule out non-blending scenario based on these centroids. We note that the face value of the centroid shift of $\delta = \sqrt{0.11^2+0.06^2}~\mathrm{pix} = 0.13~\mathrm{pix}$ can be explained by a blending star that is separated from the source by $\delta / (f_{b, I}/f_{s, I}) = 0.64~\mathrm{pix} = 166~\mathrm{mas}$. Such a small separation cannot be resolved by either OGLE or \textit{Gaia}.

\section{Model without Blending Flux in OGLE Data} % -----------------------------------------------------

The microlensing model fitted to the lightcurve points to a statistically significant blending flux. The data analyzed were obtained by two surveys. One may ask if data from both surveys indicate blending flux. Parameters of an FFP event can be constrained if photometry allows measuring the amplitude and the length of the event. For this purpose it would be best to take data at the peak and both raising and falling branches (steep part preferably) of the event. In the present case, the KMTNet data provide enough coverage. The OGLE data were taken early during the rising branch. When the OGLE observations started following night, the event was already $0.14~\mathrm{d}$ after the peak. The next night OGLE observations started when the event was almost at the baseline. We try to fit the event using OGLE data alone and found extremely wide posterior, which indicates that these data do not have time coverage that provides enough information to constrain event parameters, including blending flux. To sum up, the blending flux is constrained by the KMTNet photometry. Our analysis indicates that the event is most likely blended, but we cannot fully exclude non-blending model. 

For the sake of completeness, in this section we provide the results of the analysis with fixed zero blending in the OGLE $I$ band data. For clarity, we prime all the quantities obtained in this procedure. The microlensing model shows a significant difference: the continuous degeneracy in $u_0$ seen in Fig.~2 is broken into discrete degeneracy: $u_0\approx -0.4$ or $u_0\approx 0.4$. These solutions differ only in $u_0$ sign and we report only the positive $u_0$ case. The $\chi^2$ is worse by only 6.03 than for the free-blending case. The posterior statistics are presented in Tab.~2.

The source properties and physical parameters are derived using the same method as in Section~4:
\begin{align*}
\begin{split}
I'_{s,0} ={}& 14.019\pm0.029~\mathrm{mag}, \\ 
(V-I)'_{s,0} ={}& 0.890\pm0.025~\mathrm{mag}, \\
K'_{s,0} ={}& 12.919\pm0.059~\mathrm{mag}, \\
(V-K)'_{s,0} ={}& 1.989\pm0.077~\mathrm{mag}, \\
\theta'_\star ={}& 6.00\pm0.21~\mathrm{\mu as}, \\
\theta'_\mathrm{E} ={}& 4.89\pm0.18~\mathrm{\mu as}, \\
\mu'_\mathrm{rel} ={}& 4.24\pm0.15~\mathrm{mas/yr}.
\end{split}
\end{align*}
The physical quantities reported above do not differ significantly from those presented in Section~4.

\MakeTable{l|r}{9cm}{Same as Table~1 for model without blending in the OGLE data}
{\hline
parameter &  value\\
\hline
$t'_0$ & $2460433.5737\pm0.0015$ \\
$u'_0$ & $0.381\pm0.025$ \\
$t'_\mathrm{E}$ (d) & $0.4215\pm0.0022$ \\
$t'_\star$ (d) & $0.5172\pm0.0016$ \\
\hline
$\rho'$ & $1.227\pm0.007$ \\
\hline
$\left(\chi^2/\mathrm{dof}\right)'$ & 5407.51/5881 \\
\hline
}

\section{Discussion} % -----------------------------------------------------

We presented the discovery of a short-timescale microlensing event. The Einstein timescale of less than half a day and the Einstein ring radius of $5~\mathrm{\mu as}$ place the lens in the FFP regime. The light curve does not show signs of microlensing from a putative planet host. At the same time, the fitted microlensing model indicates significant blending flux. If blending flux is confirmed, then \event{} will be the only FFP with unresolved blending flux. If the blending flux comes from a single star then it is either a late-type dwarf in the Galactic disk or a low-luminosity giant in the Galactic bulge. The blending light can come from a stellar host of the lens, a companion to the source star, or an unrelated star. The last interpretation is unlikely because of low spatial density of stars of similar brightness in the vicinity of the event. 

The \textit{Gaia} DR3 proper motion and parallax do not allow one to constrain whether the blending light originates from the disk or the bulge. We may expect that the final \textit{Gaia} measurements will be more accurate by a factor of $\approx 2$, which most likely would still not be enough to decisively identify the blend as a disk or a bulge star.

If the blending light comes from the source companion, then spectra of the target object should reveal two stars with changing radial velocities (RVs). One can try to rule out the origin of blending light from the source companion but proving lack of RV changes will require a long-term observing campaign because the period of putative source binary is not constrained.

The origin of the blending light could be found using optical interferometry. Currently, interferometric observations of such a faint target a few mas away from a brighter star are possible only using the recently upgraded GRAVITY+ instrument on the Very Large Telescope Interferometer (VLTI; Gravity+ Collaboration \etal 2022). The interferometric observations require a nearby very bright star for fringe tracking. In the case of \event, there is a suitable star 14\farcs3 away with the brightness of 
$RP = 12.2~\mathrm{mag}$ (\textit{Gaia} DR3 4062325483339398656) and 
$K = 8.9~\mathrm{mag}$ (2MASS J17583712-2923194; Skrutskie \etal 2006). 
We note that a previous incarnation of that instrument (called GRAVITY Wide) was used for microlensing research: $\theta_\mathrm{E}$ was measured with subpercent precision (Mróz \etal 2025). GRAVITY+ is going to start observations in early 2026 and by that time the lens and the source should separate by $7~\mathrm{mas}$, i.e., well above GRAVITY+ resolution. 

In order to verify feasibility of GRAVITY+ resolving the source and the blend, we estimate their $K$-band brightness. The blend properties measured from the microlensing model:
$$I_b = 17.42^{+0.27}_{-0.17}~\mathrm{mag},
\qquad
(V-I)_b = 2.53^{+0.24}_{-0.10}~\mathrm{mag}$$
are corrected for extinction and reddening using $A_I$ and $E(V-I)$ estimated in Sec.~4:    
$$I_{b,0} = 15.98^{+0.27}_{-0.17}~\mathrm{mag},
\qquad
(V-I)_{b,0} = 1.46^{+0.24}_{-0.10}~\mathrm{mag}.$$
We then transform $(V-I)_{b,0}$ to $(V-K)_{b,0}$ (Bessell and Brett 1988):
$$(V-K)_{b,0} = 3.19^{+0.53}_{-0.19}~\mathrm{mag}$$
and calculate 
$$K_{b,0} = (V-I)_{b,0} + I_{b,0} - (V-K)_{b,0} = 14.220^{+0.062}_{-0.077}~\mathrm{mag}.$$
Finally, we add extinction: 
$$K_b = K_{b,0} + A_K = 14.433^{+0.069}_{-0.082}~\mathrm{mag}.$$
The above calculations depend on two assumptions. First, the color transformations are for giants. Using the transformation for dwarfs results in $K_b$ fainter by $0.045~\mathrm{mag}$ and with uncertainties twice larger. Second, we assume that the blend has the same extinction and reddening as derived for the bulge. If we reduce all extinction and reddening values by half, then the resulting $K_b$ is fainter by $0.16~\mathrm{mag}$. The $K_b$ combined with $K_s$ estimated earlier indicates a predicted blend-to-source $K$-band flux ratio of $0.356^{+0.044}_{-0.032}$. These predictions are presented in Fig.~6. We note that the combined $K$-band flux of source and blend corresponds to 
$12.983^{+0.056}_{-0.063}~\mathrm{mag}$.
This prediction is in a very good agreement with the VIRAC2 brightness of the target: $12.928\pm0.046~\mathrm{mag}$.
The VIRAC2 brightness differs more from the source brightness predicted for the model without blending flux in the OGLE data: $K'_s = 13.134\pm0.059~\mathrm{mag}$.

\begin{figure}[htb]
\includegraphics[width=\textwidth]{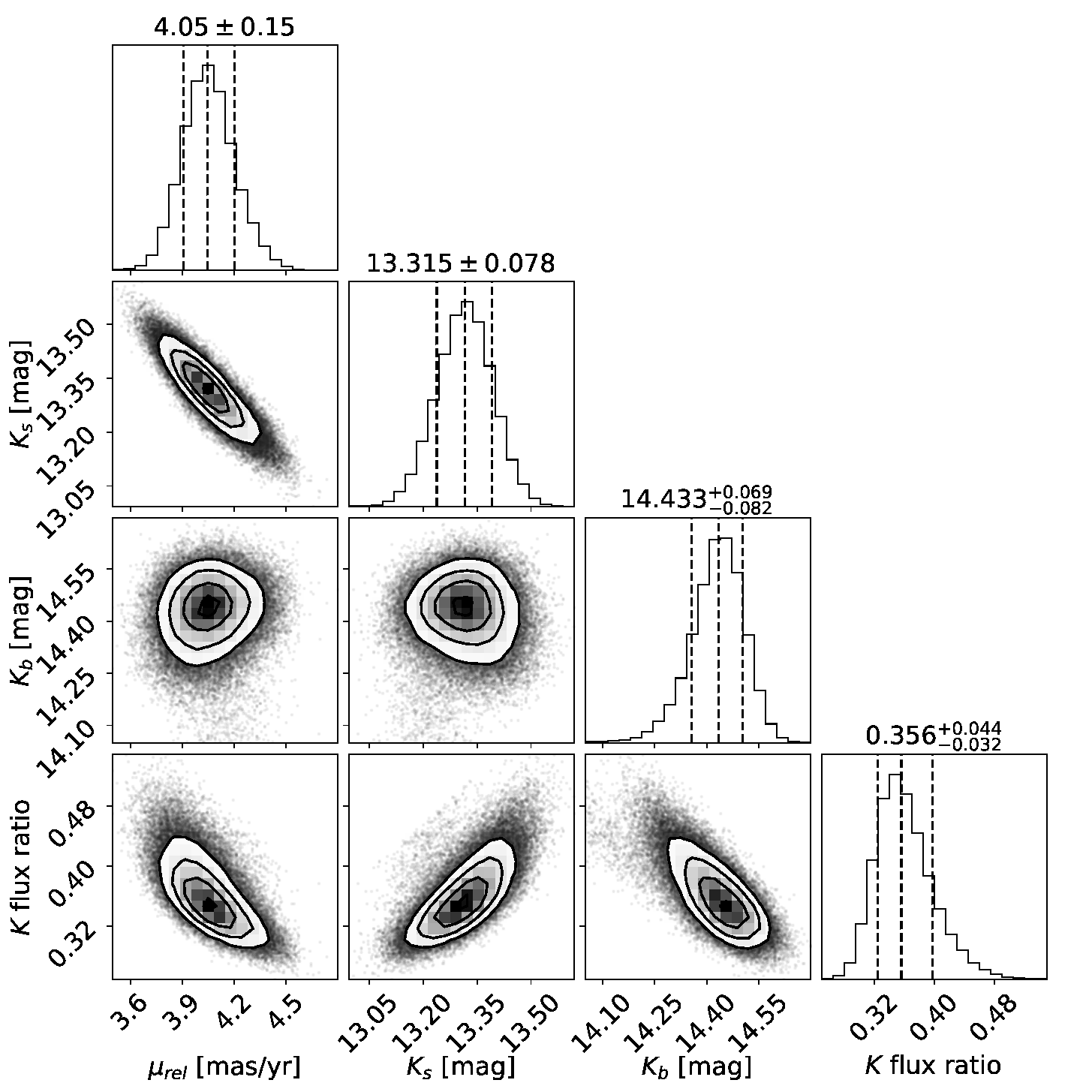}
\FigCap{Predictions for interferometric observations: relative lens-source proper motion, source brightness, lens brightness, and blend-to-source flux ration (last three for the $K$ band).}
\end{figure}

We expect that the first GRAVITY+ observation will show two stars (there is a low probability that the two stars will be unresolved). This observation will allow to constrain the $K$-band flux ratio of the two stars, which will allow improving precision of the microlensing model. The second epoch taken a year or so later will allow measuring relative proper motion of the two stars. If the relative proper motion is inconsistent with the microlensing estimation presented above, then the interpretation of the blending light as a planet host will be ruled out.

\Acknow{
% KMT
This research has made use of the KMTNet system
operated by the Korea Astronomy and Space Science Institute
(KASI) at three host sites of CTIO in Chile, SAAO in South
Africa, and SSO in Australia. Data transfer from the host site to
KASI was supported by the Korea Research Environment
Open NETwork (KREONET). This research was supported by KASI
under the R\&D program (project No. 2025-1-830-05) supervised
by the Ministry of Science and ICT.
% OGLE
The OGLE project has received funding from the Polish National Science Centre grant OPUS 2024/55/B/ST9/00447 awarded to A.U.
% Individual acknowledgements
RP acknowledges support by the Polish National Agency for Academic Exchange grant "Polish Returns 2019."
% KMT Individual acknowledgements
H.Y. and W.Z. acknowledge support by the National Natural Science Foundation of China (Grant No. 12133005). 
H.Y. acknowledges support by the China Postdoctoral Science Foundation (No. 2024M762938). 
J.C.Y. and I.-G.S. acknowledge support from U.S. NSF Grant No. AST-2108414. 
Work by C.H. was supported by the grants of National Research Foundation of Korea (2019R1A2C2085965 and 2020R1A4A2002885).
% Gaia
This work has made use of data from the European Space Agency (ESA) mission \textit{Gaia}\footnote{\texttt{https://www.cosmos.esa.int/gaia}}, processed by the \textit{Gaia} Data Processing and Analysis Consortium (DPAC\footnote{\texttt{https://www.cosmos.esa.int/web/gaia/dpac/consortium}}). Funding for the DPAC has been provided by national institutions, in particular the institutions participating in the \textit{Gaia} Multilateral Agreement.
}

\end{document}